\documentclass[prl,aps,floatfix,tightenlines,twocolumn,superscriptaddress,amsmath,amssymb,letterpaper]{revtex4-1}

\usepackage{graphicx}
\usepackage{dcolumn}
\usepackage{bm}
\usepackage{epsfig}
\def\comment#1{}

\begin{document}

\date{}

\title{Quantum communication without the necessity of quantum memories}
\author{W. J. Munro}\email{bill.munro@me.com}
\affiliation{NTT Basic Research Laboratories, NTT Corporation, 3-1 Morinosato-Wakamiya, Atsugi, Kanagawa 243-0198, Japan}
\affiliation{National Institute of Informatics, 2-1-2 Hitotsubashi, Chiyoda-ku, Tokyo 101-8430, Japan}

\author{A. M. Stephens}
\affiliation{National Institute of Informatics, 2-1-2 Hitotsubashi, Chiyoda-ku, Tokyo 101-8430, Japan}

\author{S. J. Devitt}
\affiliation{National Institute of Informatics, 2-1-2 Hitotsubashi, Chiyoda-ku, Tokyo 101-8430, Japan}

\author{K. A. Harrison}
\affiliation{Kismet Institute, Chepstow, Monmouthshire, UK} 

\author{Kae Nemoto}
\affiliation{National Institute of Informatics, 2-1-2 Hitotsubashi, Chiyoda-ku, Tokyo 101-8430, Japan}

\begin{abstract}
Quantum physics is known to allow for completely new ways to create, manipulate and store information. Quantum communication - the ability to transmit quantum information - is a primitive necessary  for any quantum internet. At its core, quantum communication generally requires the formation of entangled links between remote locations. The performance of these links is limited by the classical signaling time between such locations - necessitating the need for long lived quantum memories. Here we present the design of a communications network which neither requires the establishment of entanglement between remote locations nor the use of long-lived quantum memories. The rate at which quantum data can be transmitted along the network is only limited by the time required to perform efficient local gate operations. Our scheme thus potentially provides higher communications rates than previously thought possible.
\end{abstract}

\date{\today}

\maketitle

Quantum communication is a primitive necessary for any future quantum internet\cite{Bennett84,Gisin2002,Gisin2007,Nei00,kimble2008}---irrespective of whether such communication is over distances of centimeters or thousands of kilometers. It is more than just using a quantum channel to establish shared classical key material\cite{Gisin2002} between remote locations. Instead, for instance, it can also be used to transfer information between remote quantum computers\cite{kimble2008,bennett93}. There are a number of ways to reliably distribute quantum information between remote locations (nodes)\cite{Nei00}. The most familiar and defacto approach is to establish entanglement between the nodes and then use quantum teleportation to transfer the information from one node to the other\cite{Bennett96}. Depending on the distance between the nodes one may need quantum repeaters\cite{enk98,briegel98,dur98} at intermediate locations, whose role is to mediate entanglement between the endmost nodes. After establishing entanglement between adjacent nodes, one can use entanglement swapping\cite{briegel98} to extend the range of entanglement across the entire network of quantum repeaters\cite{Bennett96,enk98,briegel98,dur98,duan01,loock06,chen07a,sheng08,goebel08,simon07,tittel08,sangouard09,pan01,dur07}. Then standard quantum teleportation\cite{bennett93} allows the information to be moved between the endmost nodes. However the performance of such an approach is ultimately limited by the time it takes to establish the intermediary entangled links\cite{munro10}. At best this is the signaling time between adjacent nodes, but with most schemes it scales as the multiples of the round-trip time across the entire network\cite{dur07,sangouard10}. This necessitates quantum memories capable of storing information for milliseconds or longer\cite{dur07,sangouard10,munro10,collins07}. 

In this paper we present an alternative approach to quantum communication based on directly transmitting quantum information, in encoded form, across a network\cite{knill96}. As our scheme does not involve teleportation, it does not require the establishment of entangled links between nodes or long-lived quantum memories. Furthermore, as our scheme utilizes an error-correction code\cite{Knill2,kok2007,Hayes1,Ralph1} that can tolerate photon loss in excess of 50\% in the quantum channels between nodes, it allows for nodes to be spaced further apart than conventionally thought. Our scheme, in principle, allows for communication rates that significantly exceed those of existing entanglement-based schemes, yet is modest in its use of resources and is simple enough to be viable in the medium term.



\begin{figure*}
\begin{center}
\includegraphics[scale=1]{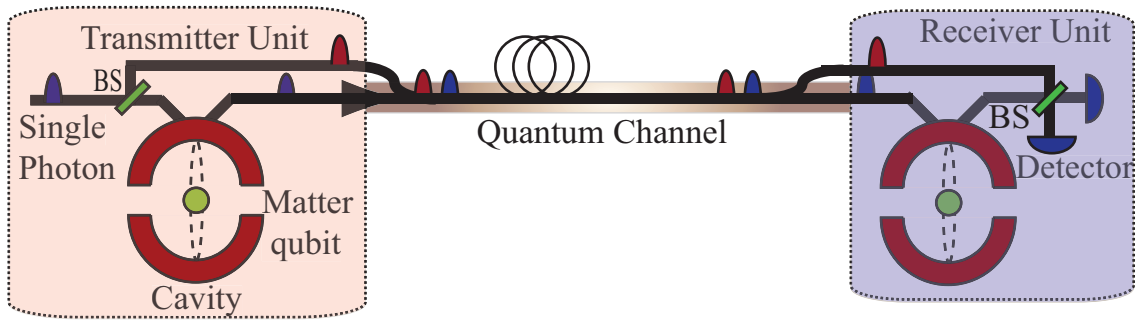}
\end{center}
\vspace{-10pt}
\caption{Schematic representation of the direct transmission of the information from one communication node to another using a transmitter unit and receiver unit, where each comprises a matter-based § contained in a cavity coupled to an optical fiber. The interaction between the matter qubit and photon is given by  $U=\exp \left[ - i \chi t \;n_a \otimes |e\rangle\langle e | \right]$ where $\chi$ is the coupling strength, $t$ the interaction time, $n_a$ the photon field number operator and $|e\rangle$ ($|g\rangle$) the excited (ground) state of the matter qubit. Choosing $\chi t$ appropriately, the photon acquires a $\pi$ phase shift only if the matter qubit is $|e\rangle$. The transfer begins by preparing the photon in the dual rail state $|0\rangle |1\rangle + |1\rangle |0 \rangle$. The first mode interacts with the matter qubit initialized as $\alpha |g\rangle+ \beta |e\rangle$ giving $\alpha |g\rangle \left(|0\rangle |1\rangle + |1\rangle |0 \rangle\right)+\beta |e\rangle \left(|0\rangle |1\rangle -  |1\rangle |0 \rangle\right)$. Measuring the matter qubit in the $|g\rangle \pm  |e\rangle$  basis projects the photonic state to $\alpha \left(|0\rangle |1\rangle + |1\rangle |0 \rangle\right)+\beta \left(|0\rangle |1\rangle -  |1\rangle |0 \rangle\right)$ (up to a known phase correction). This is then sent over the quantum channel temporally multiplexed with a heralding signal to the receiver unit which prepares the matter qubit as $|g\rangle+ |e\rangle$ and then interacts with the first photonic mode. The two modes of the photonic state are recombined on a 50/50 beamsplitter and then measured at one of the two detectors projecting the matter qubit into the state $\alpha |g\rangle+ \beta |e\rangle$ (again up to a known phase correction).} 
\label{link}
\end{figure*}

It is important to begin with a discussion of the fundamental building blocks of our scheme. We are going to consider a transmitter-receiver model\cite{munro10} [depicted and described  in Fig.~(\ref{link})]. Both the transmitter and receiver units comprise a matter qubit (an electron spin for example) located in a cavity. The matter qubits do not require a coherence time anywhere near the time required for photons to propagate between nodes. Hence our matter qubits will not be thought of as quantum memories,  instead as information processing qubits. The transmitter unit contains a single photon source while the receiver unit contains a single photon detector. Coupling the transmitter unit in one node to a receiver in another node via an optical fiber then allows a quantum state to be directly transmitted between those nodes. 

This would seem to be an ideal solution but a serious problem exists. Channel  and coupling losses (as well as source and detector inefficiencies)  will degrade the quantum state that is being transmitted\cite{sangouard10}. This is may not be an issue for quantum key distribution (QKD) applications based on BB84 as the key material can be post selected on only those that arrive at the end node\cite{Gisin2002,Gisin2007} but at a significant expense on the rate. Also for other communication and computational tasks it is not that simple - entanglement could be necessary here. Post selection would require long lived quantum memories at the end nodes\cite{sangouard10,munro10}. This issue can be overcome (to a certain extent) by encoding our state using a general error-correction code\cite{knill96,devitt,Got09,jiang09,Fowler1}. The error correction code could protect against the loss as quantum signals are sent between nodes. There are many codes that one can use but these generally tolerate much less than 50\% loss\cite{devitt,Got09,jiang09,Fowler1}. One particular type of code that can tolerate much more loss is the family of redundant quantum parity codes first considered in the context of linear optical quantum computation\cite{Knill2,kok2007,Hayes1,Ralph1}. Here a single photonic quantum state $\alpha |0\rangle+ \beta |1\rangle$ can be encoded as
\begin{eqnarray}
|\Psi \rangle^{(m,n)}= \alpha |+\rangle^{(m)}_1 \ldots |+\rangle^{(m)}_n + \beta  |-\rangle^{(m)}_1 \ldots |-\rangle^{(m)}_n,
\label{repeater-parity}
\end{eqnarray}
where $n$ is the number of logical qubits and $m$ is the number of physical qubits in each logical qubit. The logical qubit basis states are given by $|\pm\rangle^{(m)} \equiv |0\rangle^{\otimes m} \pm |1\rangle^{\otimes m}$. This encoding has the property that a quantum state can be recovered as long as at least one logical qubit is recovered without the loss of any of its physical qubits $and$ at least one physical qubit is retained from each of the other logical qubits\cite{Hayes1,Ralph1}.

\begin{figure*}
\begin{center}
\includegraphics[scale=0.8]{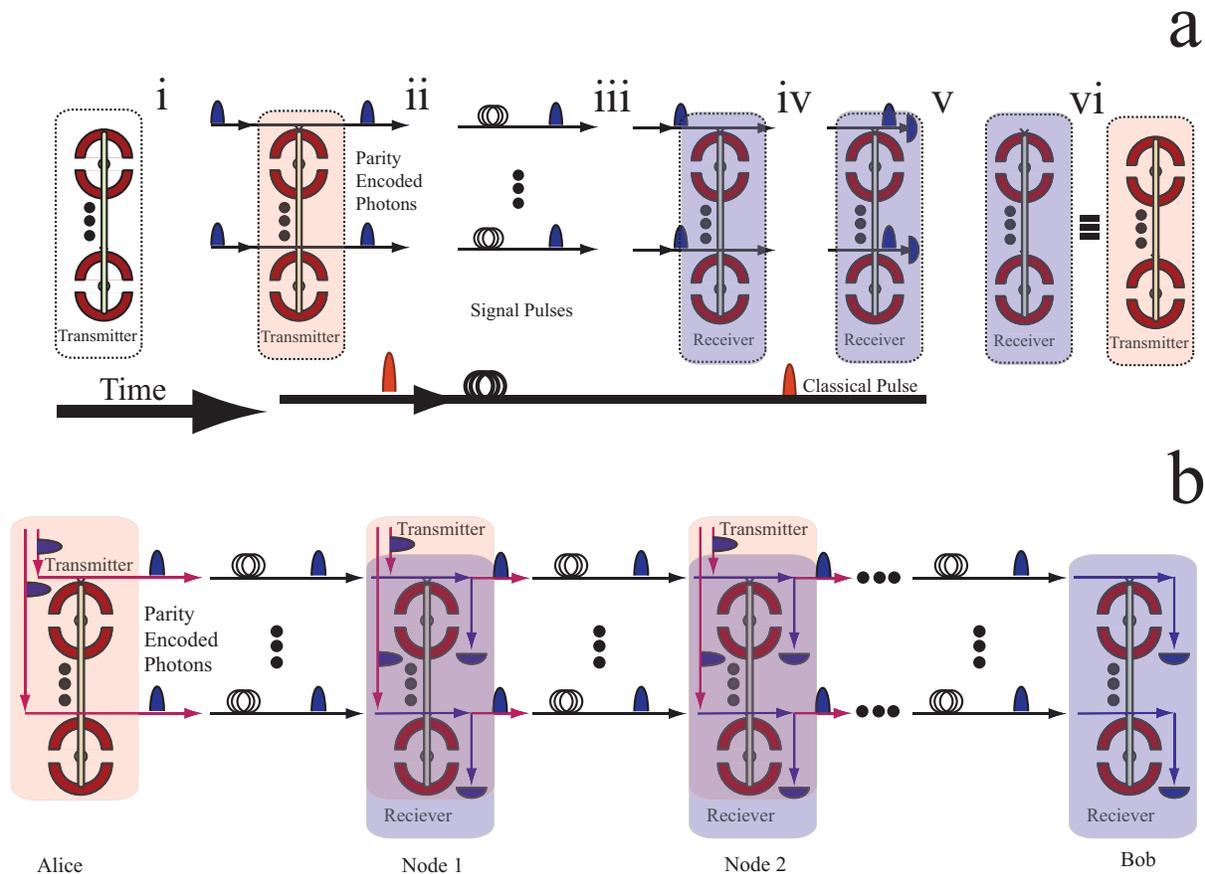}
\end{center}
\vspace{-10pt}
\caption{
Schematic representation in a of  the transmission of a quantum signal using a redundant quantum parity code. It begins in i with the construction of the redundant quantum parity code on the matter qubits and the encoding of a quantum state with parameters $\alpha,\beta$. This requires highly efficient single- and two-qubit  operations.  In ii the quantum states of each of the matter qubits are transferred to photonic states, as detailed in Fig.~(\ref{link}). The matter qubits are now free to be used again. In iii the photons are transmitted over the  lossy channel to the remote receiver units. In iv the information encoded on the photonic states is transferred to matter qubits. Next in v the photons are measured to herald loss events. The loss-affected parity blocks are then disentangled from the remaining part of the quantum state by measuring one or more of the matter qubits in the $Z$ basis. Parity blocks that have no loss are left untouched and so we have now finished the transfer of information between nodes. We can now if required use the general error correction properties inherent in the code to correct small dephasing errors. Finally in vi the quantum state at the receiver side can be  re-encoded back to its original size by using the discarded matter qubits. The receiver can then act as a transmitter to send the signal to the next node and so the state can be sent over long distances as shown in b.} 
\label{linka}
\end{figure*}


The redundant quantum parity code  can then be used in a repeater scheme to transmit signals between nodes as is depicted and described  in Fig.~(\ref{linka}a). The transmitter unit moves information from the matter qubits to photonic ones while the receiver units operate in reverse to transfer the information from the photonic ones back to matter qubits where the parity code can be checked and errors corrected. After re-encoding back to the full size the matter qubits can then be used in the transmitter stage to send photons to the next communications node and so on. This in principle allows the transmission of the original quantum state over long ranges\cite{classical,classical2} (shown in Fig.~(\ref{linka}b)). How do we know if we have been successful in transmitting our quantum signal between nodes? This occurs if two primary requirements are met: First, at least one photon must arrive per logical qubit. The probability of no photon arriving in the m photon logical qubit is $p_f=\left(1-p\right)^m$ where $p$ is the probability of a single photon arriving at the adjacent node. Second, at least one of the logical qubits must arrive with no loss. With $n$ logical qubits the probability we do not receive at least one logical qubit without error is $p_f=1-[1-(1-p)^m]^n+[1-p^m-(1-p)^m]^n$. Once we set our target failure error rates, $p_f$, and given $p$, we can determine the minimum $n$ and $m$ required. For illustrative purposes Table (\ref{table:resources}) shows estimates of $m$ and $n$ for various $p$. The resources are quite modest for $p>0.60$ but increase significantly as $p$ decreases. However, this can be overcome by realizing that a single photon can carry more that one qubit of information (photons have multiple degrees of freedom, such as time bin, polarization, spatial modes)\cite{Popescu,Mair01,Thew04,Howel05,Oemrawsingh05,Barreiro05,Kwiat97}.  This means fewer photons need to be send through the channel and so we have a higher chance of receiving them all. 

Encoding more than one qubits on a photon has long being known in the linear optical quantum computation community as a way to reduce the number of photons required to undertake certain tasks. It can be used to the same effect here in quantum communication and networks. Consider a general two matter qubit state $\alpha_0 |gg\rangle+\alpha_1 |ge\rangle+\alpha_2 |eg\rangle+\alpha_3 |ee\rangle$ which we want to transfer on to a single photon prepared into a equal superposition state over four spatial modes $ |10\rangle|00\rangle+|01\rangle|00\rangle+|00\rangle|10\rangle +|00\rangle|01\rangle$. The transfer is achieved by performing a CPhase operation between the first matter qubit and second photonic mode and a CPhase gate between the second matter qubit and fourth photonic mode. Measuring the matter qubits in the $|g\rangle\pm |e\rangle$ basis gives the spatial encoded photonic state
\begin{eqnarray}
&& \left\{\alpha_0 \pm \alpha_1\pm \alpha_2+\alpha_3 \right\} |10\rangle|00\rangle \nonumber \\
 &+& \left\{\alpha_0 \pm \alpha_1\mp\alpha_2-\alpha_3 \right\} |01\rangle|00\rangle \nonumber \\
&+&\left\{\alpha_0 \pm \alpha_1\pm \alpha_2+\alpha_3 \right\} |00\rangle|10\rangle \nonumber \\
&+& \left\{\alpha_0 \pm \alpha_1\mp \alpha_2-\alpha_3 \right\} |00\rangle|01\rangle \nonumber
\end{eqnarray}
The four spatial photonic modes can then be send temporally multiplexed across the channel to the receiver units where a similar procedure can be used to transfer it back to matter qubits. 

Now we encode our redundancy parity code uniformly over all the photons and degrees of freedom such that one photon does not carry more than one qubit from a particular parity block\cite{special2}.  While this may seem trivial, it significantly reduces the number of physical qubits required within each communication node, as is shown in Table (\ref{table:resources}).

\begin{table}
\vspace{2pt}
\centering 
\begin{tabular}{c c c c c c c c} 
\hline\hline 
$p$ & $m$ &  $n$ &  $R_G$ (Hz) &  $m'$ &   $n'$ &  $R'_G$ (Hz) \\
\hline
0.95 & 3 & 4  & $10^7$ &  3 &  5 &  $10^7$ \\ 
0.90 & 4 & 8  & $10^7$ &  3 &  6 &  $10^7$ \\ 
0.82 & 6 & 22  & $10^7$ &  4 &  10 &  $10^7$ \\ 
0.67& 13 & 1500   & $10^7$ &  6 & 18 & $10^7$\\
0.60 &  &   & $10^7$ &  8 &  25 &  $10^7$ \\ 
0.55 &  &   & $10^7$ &  9 &  31 &  $10^7$ \\ 
\hline
\end{tabular}
\caption{Matter qubit resources $m$, $n$ ($m'$, $n'$) required for the redundant quantum parity code with one qubit/photon (multiple qubits/photon) for various  $p$,  the probability of a single link being successfully established between adjacent nodes. $m,m'$ represent the number of qubits with the block. The resources are calculated for a maximum failure probability $p_f \sim<  0.001$. Also given are the rates $R_G$ and $R'_G$ at which quantum states can be transferred between adjacent nodes  as a function of p. }
\label{table:resources}
\end{table}

So far we have focussed on loss as our main source of error. However in realistic systems we also have gate errors that occur during the encoding and decoding circuits or during the  transferral between matter-based and photons.  We also have measurement errors as each logical qubit with photon loss needs to be measured out in the Z basis. Imperfect Z measurements can lead to a logical error on our encoded state, however this can be minimized by using a majority voting tactic. It is likely that more than one qubit will have been successfully transferred within each loss affected logical qubit and so by measuring them all in the Z basis we can decrease the effect of such errors.  While not designed to handle general errors the redundant quantum parity code has some ability to correct these as the code is based on a generalization of the Shor code\cite{sho95} for $n,m\geq 3$. In the situation where $q$ logical qubits are received in tact we can correct  $\frac{m-1}{2}$ bit flip errors within each logical qubit and also $\frac{q-1}{2}$ sign flip errors.  Without additional error correction on top of the parity code the local error rate from the gates and measurements sets a limit on how much loss our code can tolerate and thus how far we can transmit our quantum signal.  


The next question is what is the performance of our system and how does it compare to others in the literature? Initially assuming the local operations are good enough so that we do not require local purification it is straightforward to determine the performance. In Table (\ref{table:resources}) we show the communication rate for various $p$ where we have assumed that all local gate operations can be done with a total time of 100ns implying a raw communication rate of  $10^7$ quantum states being transmitted per second.  With $p=0.60$ we require at least 200 matter qubits per node. However what distance does this correspond to? The probability $p$ is given by $p= p_s p_d p_c^2 \exp \left[-{\cal L}/{\cal L}_0\right]$ where $p_s$ is the probability the source emits a single photon, $p_d$ the probability a single photon is detected and $p_c$ the probability of coupling a photon to the cavity. $\cal L$ is the distance between the adjacent nodes and ${\cal L}_0$ is the attenuation length of the channel (25km for commercial fiber). If we choose $ p_s=p_d=p_c=0.97$ then $p=0.60$ corresponds to  a distance between nodes of ${\cal L}=10$km. Having 80 communication nodes would allow the transmission of quantum states over 800km with a success probability exceeding 98\% using $80\times 200$ matter qubits ($R'_G /{\rm total\;matter\;qubits}\sim  625$). If the transfer fidelity between adjacent nodes is of the order of  $99.9\%$ then an overall fidelity of 90\% could be achieved without the need for further error correction (or purification).  This seems to indicate a good potential  which could be improved further with faster local operations but this does not give us an indication of whether this performance is good or bad.  To do this we need to make a comparison to other known schemes.


There are a number of other schemes we can use as a comparison\cite{Gisin2002,Gisin2007,loock06,dur07,sangouard10,munro10,jiang09,Bennett96,enk98,briegel98,dur98,duan01,chen07a,sheng08,goebel08,simon07,tittel08,sangouard09,pan01}. The most obvious is the direct transmission of the photon\cite{Gisin2007,sangouard10} over the entire 800km which occurs with a success probability $p\sim 10^{-15}$ using two qubits (which would correspond to a rate $<\sim 10^{-6}$Hz). The original DLCZ protocol\cite{duan01} gives a rate of $10^{-3}$Hz using 32 qubits (atomic ensembles) divided between 16 links and achieving a fidelity of 90\%. An improved protocol\cite{simon07} based on photon-pair sources and multimode memories  (able to store 100 modes) to implement a temporally multiplexed version of the DLCZ protocol achieves a rate of $0.1$Hz using 16 links.  Alternatively a spatially multiplexed scheme\cite{munro10} with the 20 links (with 50 qubits/node) separated by 40km and deterministic local gates  can achieve an approximate rate of $2400$Hz.  To compare these we could look at the rate of data transfer over 800km divided by the total number of qubits used. In such a case our scheme presented here is nearly two orders of magnitude better. These considerations have assumed that the number of links is low enough such that purification is not likely to be required.  With purification\cite{loock06}  it was found that  a rate of $80$Hz could be achieved over the same distance using 32 links each composed of 16 qubits.  Using a fully error corrected scheme\cite{jiang09} it was found a rate of $100$Hz could be achieved using 80 links each composed of $30-150$ qubits. So in terms of the rate divided by the total number of qubits over the whole network these schemes can achieve $R/{\rm total\;qubits}\sim  10^{-4} - 1$. Our scheme is between two to four orders of magnitude better. 

\begin{figure*}
\begin{center}
\includegraphics[scale=0.8]{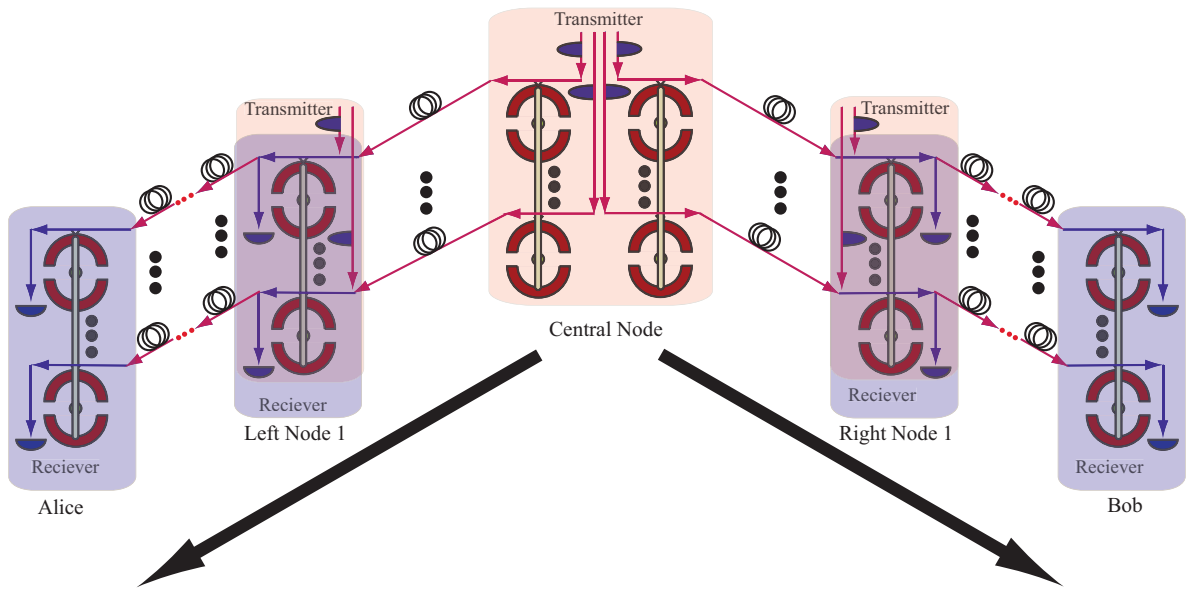}
\end{center}
\vspace{-10pt}
\caption{Schematic representation illustrating how the direct transmission scheme can be used in a butterfly arrangement\cite{munro10} to distribute entanglement between Alice and Bob without the need for long-lived quantum memories.} 
\label{entangle}
\end{figure*}

Finally while our approach has so  far been about sending information along the network, it can also be used to generate entanglement between the end nodes as we depict in Fig(\ref{entangle}) still without the need for long-lived quantum memories. In this situation a parity encoded Bell state is created at the central node. One half is send to the left and the other half to the right using the redundant parity code on each. This entanglement could be used, for instance, in device-independent QKD and Bell tests over long distances.

To summarize, we have presented a quantum communication scheme based on parity codes\cite{newcodes}, which allows the direct transmission of quantum states between distributed systems. The scheme gives high communication rates with modest resource usage for modestly separated nodes (less than 10 km). More importantly our scheme does not require the use of long-lived quantum memories but does requires accurate local gates. This potentially enables communication rates several orders of magnitude faster than current schemes. This work shows the importance of efficient interfaces between light and matter.  While we have focused on sending information directly it is also straightforward to consider a butterfly arrangement\cite{munro10} to distribute entanglement between the end nodes  (still without the need for long-lived memories).  Our scheme has important implications for the development of quantum communication technologies in the future. Without special attention to being paid to the system design one will easily become limited by classical signalling costs. 

{\it Acknowledgements}: We thank Norbert Lutkenhaus and Liang Jiang for valuable discussions and pointing out a missing term in one of the probability distributions. This work was supported in part  by the Japan Society for the Promotion of Science (JSPS) through its Funding Program for World-Leading Innovative R\&D on Science and Technology (FIRST Program) and a grant from the National Institute of Information and Communications Technology (NICT).

{\it Note added:} The numbers in Table I have changed a little from the published version due to the modification to add an extra term to the probability distribution. This restricts the total loss to less than 50 percent.


\begin{thebibliography}{99}

%
\bibitem{Bennett84} 
Bennett, C. H. \& Brassard, G. Quantum cryptography: public key distribution and coin tossing, in Proceedings of IEEE International Conference on Computers, Systems, and Signal Processing, 175-179 (IEEE Press, 1984).
%
\bibitem{Gisin2002} 
Gisin, N., Ribordy, G., Tittel, W. \& Zbinden, H. Quantum cryptography. \textit{Rev. Mod. Phys.} {\bf 74}, 145-195 (2002).
%
\bibitem{Gisin2007} 
Gisin, N.  \& Thew, R. Quantum Communication.  \textit{Nature Photon.} {\bf 1}, 165 - 171 (2007)  
%
\bibitem{Nei00} 
Nielsen, M. A. \& Chuang, I. L. \textit{Quantum Computation and Quantum Information} (Cambridge Univ. Press, 2000).
%
\bibitem{kimble2008}  
Kimble, H. J. The quantum internet.   \textit{Nature} {\bf 453}, 1023 - 1030 (2008).
%
\bibitem{bennett93} 
Bennett, C. H., Brassard, G., Crepeau, C., Jozsa, R., Peres, A. \& Wootters, W. K.  Teleporting an unknown quantum state via dual classical and Einstein-Podolsky-Rosen channels. \textit{Phys. Rev. Lett.} {\bf 70}, 1895 - 1899 (1993).
%
\bibitem{Bennett96} 
Bennett, C. H.,  Brassard, G., Popescu, S., Schumacher, B., Smolin, J. A., \& Wootters, W. K. Purification of Noisy Entanglement and Faithful Teleportation via Noisy Channels.  \textit{Phys. Rev. Lett.} {\bf 76}, 722 - 726 (1996).
%
\bibitem{enk98} 
Enk, S. J., Cirac, J. I. \& Zoller, P. Photonic channels for quantum communication. \textit{Science} {\bf 279}, 205-208 (1998).
 %
\bibitem{briegel98} 
Briegel, H.-J., D\"ur, W., Cirac, J. I. \& Zoller, P. Quantum repeaters: the role of imperfect local operations in quantum communication. \textit{Phys. Rev. Lett.} {\bf 81}, 5932-5935 (1998).
%
\bibitem{dur98} 
D\"ur, W., Briegel, H.-J., Cirac, J. I. \& Zoller, P.  Quantum repeaters based on entanglement purification. \textit{Phys. Rev. A} {\bf 59}, 169-181 (1999).
%
\bibitem{duan01}
Duan , L. M., Lukin, M. D., Cirac, J. I. \& Zoller, P. Long-distance quantum communication with atomic ensembles and linear optics. \textit{Nature} {\bf 414}, 413-418 (2001).
%
\bibitem{loock06}
Van Loock, P., Ladd, T. D., Sanaka, K., Yamaguchi, F., Nemoto, K., Munro, W. J. \& Yamamoto, Y. Hybrid quantum repeater using bright coherent light. \textit{Phys. Rev. Lett.} {\bf 96}, 240501 (2006).
%
\bibitem{chen07a}
Zhao, B., Chen, Z.-B., Chen, Y.-A., Schmiedmayer, J. \& Pan, J.-W. Robust creation of entanglement between remote memory qubits. \textit{Phys. Rev. Lett.} {\bf 98}, 240502 (2007).
%
\bibitem{sheng08}
Yuan, Z., Chen, Y., Zhao, B., Chen, S., Schmiedmayer. J.  \& Pan, J.-W.  Experimental demonstration of a BDCZ quantum repeater node. \textit{Nature} {\bf 454}, 1098-1101 (2008). 
%
\bibitem{goebel08}
Goebel, A. M., Wagenknecht, G., Zhang, Q., Chen, Y., Chen, K., Schmiedmayer, J.  \& Pan, J.-W. Multistage Entanglement Swapping.  \textit{Phys. Rev. Lett.} {\bf 101}, 080403 (2008).
%
\bibitem{simon07}
Simon C., de Riedmatten H., Afzelius M., Sangouard N., Zbinden H. \& Gisin N.  Quantum Repeaters with Photon Pair Sources and Multimode Memories.  \textit{Phys. Rev. Lett} {\bf 98}, 190503 (2007).
%
\bibitem{tittel08}
Tittel, W., Afzelius, M., Chaneli\'{e}re, T., Cone, R.L., Kr\"oll, S., Moiseev, S.A., Sellars, M. Photon-echo quantum memory in solid state systems. \textit{Laser Photon. Rev.} {\bf 4}, 244 - 267 (2009).
%
\bibitem{sangouard09}
Sangouard, N., Dubessy, R., \& Simon, C. Quantum repeaters based on single trapped ions.  \textit{Phys. Rev. A} {\bf 79}, 042340 (2009).
%
\bibitem{pan01}
Pan, J.-W., Simon, S., Brukner, C. \& Zeilinger, A. Entanglement purification for quantum communication. \textit{Nature} {\bf 410}, 1067-1070 (2001).
%
\bibitem{dur07} 
D\"ur, W. \& Briegel, H. J. Entanglement purification and quantum error correction. \textit{Rep. Prog. Phys.} {\bf 70}, 1381-1424 (2007).
%
\bibitem{munro10}
Munro, W. J., Harrison, K.A., Stephens, A.M., Devitt, S.J. \& Nemoto, K. From quantum multiplexing to high-performance quantum networking. \textit{Nature Photon.} {\bf 4} , 792-796 (2010). 
%
\bibitem{sangouard10} Sangouard, N., Simon, C., de Riedmatten, H. \& Gisin, N. Quantum repeaters based on atomic ensembles and linear optics. {\it Rev. Mod. Phys.} {\bf  83}, 33 - 80 (2011) and references within.
%
\bibitem{collins07} 
Collins, O. A., Jenkins, S. D., Kuzmich, A. \& Kennedy, T. A.  Multiplexed Memory-Insensitive Quantum Repeaters. \textit{Phys. Rev. Lett.} {\bf  98}, 060502 (2007).
%
\bibitem{knill96} Knill, E. \& Laflamme R.  Concatenated Quantum Codes. Preprint at http://arxiv.org/abs/quant-ph/9608012 (1996).
%
\bibitem{Knill2}
Knill, E., Laflamme, R. \& Milburn, G.J. A scheme for efficient quantum computation with linear optics.  \textit{Nature} {\bf 409,} 46-52 (2001).
%
\bibitem{kok2007} Kok, P., Munro, W.J., Nemoto, K., Ralph, T.C., Dowling, J.P. \& Milburn, G.J.  Linear optical quantum computing with photonic qubits,  \textit{Rev.  Mod.  Phys.}  {\bf79} , 135-174 (2007), and references within.
%
\bibitem{Hayes1}
Hayes, A.J.F., Gilchrist, A., Myers, C.R. \& Ralph T.C.  Utilizing encoding in scalable linear optics quantum computing.  \textit{J. Opt. B.} {\bf 6,} 533-541 (2004).
%
\bibitem{Ralph1}
Ralph, T.C., Hayes, A.J.F. \& Gilchrist, A. Loss-Tolerant Optical Qubits.  \textit{Phys. Rev. Lett.} {\bf 95,} 100501 (2005).
%
\bibitem{devitt} Devitt, S.J., Nemoto, K. \& Munro, W.J.  The idiots guide to Quantum Error Correction. Preprint at http://arxiv.org/abs/0905.2794v2 (2009).
%
\bibitem{Got09} Gottesman, D. An Introduction to Quantum Error Correction and Fault-Tolerant Quantum Computation. Preprint at http://arxiv.org/abs/0904.2557 (2009).
%
\bibitem{jiang09}
Jiang, L., Taylor, J. M., Nemoto, K., Munro, W. J., Van Meter, R. \& Lukin, M. D. Quantum repeater with encoding. \textit{Phys. Rev. A} {\bf 79}, 032325 (2009).
%
\bibitem{Fowler1} Fowler, A. G., Wang, D. S., Hill, C. D., Ladd, T. D., Van Meter, R. \& Hollenberg, L. C. L. Surface code quantum communication. \textit{Phys. Rev. Lett.} {\bf  104}, 180503 (2010).
%
\bibitem{classical} 
The transfer of quantum state from one node to another is very much like the way classical communication is undertaken today.  For quantum communication it means that the node originating the message does not need to know the full path to the final receiving node. We can use a telephone exchange-like model and well-known routing protocols. It also means resources do not have to allocated for long times while messages are being sent.
%
\bibitem{classical2} 
To extend the range over longer distances we need some form of general purification or error correction due to the fact that our state will not be perfectly transmitted. It will have a finite fidelity say $F$. If we had n+1 nodes then the fidelity would drop by $F^n$.  This is still an exponential scaling (through potentially much slower than if we have just transmitted the signal unencoded through the channel). The exponential scaling can be termed into a polynomial scaling with the aid of general error correction.  Our redundant parity code has a limited ability to correct such general errors.
%
\bibitem{Popescu} 
Popescu, S., An optical method for teleportation. Preprint at http://arxiv.org/abs/quant-ph/9501020 (1995).
%
\bibitem{Mair01}  Mair, A., Vaziri, A., Weihs, G. \& Zeilinger A. Entanglement of the orbital angular momentum states of photons. \textit{Nature} {\bf 412}, 313 - 316 (2001).
%
\bibitem{Thew04}  
Thew, R.T., Acin, A., Zbinden, H. \& Gisin, N. Bell-Type Test of Energy-Time Entangled Qutrits, \textit{Phys. Rev. Lett.} {\bf 93}, 010503 (2004).
%
\bibitem{Howel05} 
OSullivan-Hale, M. N., Khan, I. A., Boyd, R.W. \& Howell, J.C. Pixel Entanglement: Experimental Realization of Optically Entangled d=3 and d=6 Qudits. \textit{Phys. Rev. Lett.} {\bf 94}, 220501 (2005).
%
\bibitem{Oemrawsingh05}  
Oemrawsingh, S. {\it et al.} Experimental Demonstration of Fractional Orbital Angular Momentum Entanglement of Two Photons. \textit{Phys. Rev. Lett.} {\bf 95}, 240501 (2005).
%
\bibitem{Barreiro05}  
Barreiro, J.T., Langford, N.K., Peters, N.A. \& Kwiat, P.G. Generation of Hyperentangled Photons Pairs. \textit{Phys. Rev. Lett.} {\bf 95}, 260501 (2005).
%
\bibitem{Kwiat97}  Kwiat, P.G. Hyper-entangled states.  \textit{J. Mod. Opt.} {\bf 44}, 2173 - 2184 (1997).
%
\bibitem{special2}
If the photon carried a logical qubits worth of information and it was lost then the entire redundant quantum parity code would be destroyed.
%
\bibitem{sho95} 
Shor, P.W.  Scheme for reducing decoherence in quantum computer memory. \textit{Phys. Rev. A} {\bf 52} R2493 - R2496, (1995).
%
\bibitem{newcodes}
Our scheme is not restricted to using  simple redundant parity codes. Instead we could use a  more general code of the form $[[m, k, 2t+1]]$ where $k$ encoded qubits are encoded into $m$ physical qubits and can correct $t$ errors. In our situation we require $k$ as possible. As long as we receive one of these encoded qubits without loss errors we can disentangle the logical qubits from our quantum state we are protecting. Next codes could be chosen with $t>0$ such that only $m-t$ qubits need to be received before we need to discard the logical qubit. This can dramatically reduce the size $n$ required for the redundancy code.


\end{thebibliography}
\end{document}